\documentclass[12pt]{article}

\usepackage{cite}
\usepackage{graphicx}
\usepackage{epsfig}
\usepackage{amsfonts}
\usepackage{amssymb}
\usepackage{amsmath,amssymb}
\usepackage{color}
\usepackage{epsf,epsfig}

\date{\today}

\newcommand{\vphi}{\varphi}

\newcommand{\ee}{\end{equation}}
\newcommand{\eea}{\end{eqnarray}}
\newcommand{\be}{\begin{equation}}
\newcommand{\bea}{\begin{eqnarray}}

\begin{document}
\begin{center}

{\Large \bf 
An analytic effective model  for hairy black holes} 
\vspace{0.6cm}
\\
{\large Yves Brihaye}$^{1}$, {\large  Thomas Delplace}$^{1}$,
{\large Carlos Herdeiro}$^{2}$, 
and
{\large Eugen Radu}$^{2}$
\vspace{0.3cm}
\\ 
$^{1}${\small Physique-Math\'ematique, Universite de
Mons-Hainaut, Mons, Belgium}
\\
$^{2}${\small Departamento de F\'\i sica da Universidade de Aveiro  and \\ Centre for Research and Development in Mathematics  and Applications (CIDMA)} \\
   {\small Campus de Santiago, 3810-183 Aveiro, Portugal}
\end{center}

\small{
\begin{abstract}Hairy black holes (BHs) have macroscopic degrees of freedom which are not associated with a Gauss law. As such, these degrees of freedom are not manifest as quasi-local quantities computed at the horizon. This suggests conceiving hairy BHs as an interacting system with two components: a ``bald" horizon coupled to a ``hairy" environment. Based on this idea we suggest an effective model for hairy BHs -- typically described by numerical solutions -- that allows computing \textit{analytically} thermodynamic and other quantities of the hairy BH in terms of a fiducial bald BH. The effective model is \textit{universal} in the sense that it is only sensitive to the fiducial BH, but not to the details of the hairy BH. Consequently, it is only valid in the vicinity of the fiducial BH limit. We discuss, quantitatively, the accuracy of the effective model for asymptotically flat BHs with synchronised hair, both in $D=4$ (including self-interactions) and $D=5$ spacetime dimensions. We also discuss the applicability of the model to synchronised BHs in $D=5$ asymptotically $AdS$ and static $D=4$ coloured BHs, exhibiting its limitations. 
\end{abstract}
}

\date{\today}
 
 

\section{Introduction}
The 1970s represent a golden era for the theoretical study of black holes (BHs). At the classical level it was understood that elctrovacuum BHs are remarkably featureless, being completely classified by a small number of \textit{macroscopic} independent degrees of freedom: mass, angular momentum and electric (possibly also magnetic) charge - see~\cite{Chrusciel:2012jk} for a review. At the quantum level, on the other hand, the visionary works of Bekenstein~\cite{Bekenstein:1973ur} and Hawking~\cite{Hawking:1974sw} heralded BHs as gateways into the realm of quantum gravity, by showing they are thermodynamical objects, and, in particular that they have an entropy, geometrically computed as the horizon area. Understanding and counting the \textit{microscopic} degrees of freedom associated to this entropy became a primary challenge for any quantum gravity candidate theory. Two decades later, some remarkable success was obtained within String Theory, werein, starting with~\cite{Strominger:1996sh,Callan:1996dv},  it was possible to identify and count the microscopic degrees of freedom that explain the classical geometric entropy, defined by the BHs macroscopic degrees of freedom, albeit only in some particular classes of BHs.

The macroscopic simplicity of electrovacuum BHs suggested the ``no-hair" conjecture~\cite{Ruffini:1971bza}: that the endpoint of gravitational collapse is a stationary BH completely described by a small set of macroscopic degrees of freedom, all of which should be associated to Gauss laws. In particular, this means such degrees of freedom are manifest at the horizon and can be computed therein as quasi-local quantities ($e.g.$ the Komar integrals~\cite{Komar:1958wp} associated to the mass and angular momentum of stationary axisymmetric BHs and the Gauss law associated to the electric charge). This, in turn, ties up nicely with the microscopic picture, and the view that the horizon contains all relevant BH information. 

The discovery of ``hairy" BHs in a variety of models (see $e.g.$~\cite{Bekenstein:1996pn,Herdeiro:2015waa,Volkov:2016ehx} for reviews) has overshadowed this conceptually simple picture. These BHs have extra macroscopic degrees of freedom not associated to a Gauss law. Therefore they do not seem to be associated to any quasi-local conserved quantity computable at the horizon level. This raises interesting questions, on how the microscopic description of the BH captures these extra macroscopic degrees of freedom, but it also suggests an effective model for obtaining an (in general) analytic approximation for physical and thermodynamical quantities of the hairy BHs associated to the horizon~\cite{Herdeiro:2017phl}. 

The basic idea of the effective model sketched in~\cite{Herdeiro:2017phl} (and suggested by the numerical evolutions in~\cite{East:2017ovw}), therein called \textit{quasi-Kerr horizon model}, is that due to the absence of further local charges at the horizon, the horizon of the hairy BH is well approximated by the horizon of a fiducial bald BH but with different parameters. In a sense, the hairy BH can be conceived as a coupled system of a bald horizon with an external ``hair" environment. Naturally, the system is interacting and the non-linearities of the underlying gravity-matter system introduce a non-trivial deformation of the ``bald" horizon. But in the feeble hair regime, when only a small percentage of the overall spacetime energy is contained in the matter field, these non-linearities are expected to be small, and the horizon should still behave as that of the bald fiducial BH, but with shifted parameters to take into account the mass and angular momenta that is no longer inside the horizon but rather in the matter environment. One could expect such simple model to yield errors in the thermodynamics quantities of the order of the deviation from the fiducial bald BH. The findings in~\cite{Herdeiro:2017phl}, however, revealed that this effective model gives an unexpectedly good approximation, sometimes with deviations of $\sim\mathcal{O}(1\%)$ even for fairly large deviations from the bald BH, $e.g.$, when $\sim\mathcal{O}(30\%)$ of the spacetime energy is stored in the matter field. 

The purpose of this paper is to investigate the applicability and accuracy of the model by considering further examples of hairy BHs. Thus, after reviewing the assumptions, basic statement and corollaries of the effective model in Section~\ref{section2}, we consider two applications in Section~\ref{section30}: we apply it to Kerr BHs with synchronised hair and self-interactions~\cite{Herdeiro:2015tia} in Section~\ref{section3} and to five dimensional ($D=5$) Myers-Perry BHs with synchronised hair~\cite{Brihaye:2014nba} in Section~\ref{section4}. In both these examples the effective model performs well. In the $D=4$ case the accuracies are comparable to those described at the end of the last paragraph. In the $D=5$ case, there is a mass gap between the hairy BHs and the fiducial BH. This means that the fiducial BH geometry is never approached globally, but only locally. In this case we find that, even for very hairy BHs, for which $\sim\mathcal{O}(90\%)$ of the spacetime energy is stored in the matter field, the model can yield errors of $\sim\mathcal{O}(1\%)$ for some physical quantities.  To exhibit also the limitations of the effective model, we consider in Section~\ref{section5} two further applications: to the $D=5$ $AdS$ Myers-Perry BHs with synchronised hair~\cite{Dias:2011at} and to the coloured BHs in Einstein-Yang-Mills theory~\cite{Volkov:1989fi}. With these applications, we illustrate either difficulties in the formalism, or unimpressive accuracies. In Section~\ref{section6} we present some final remarks, in particular speculating about the underlying reason for the good accuracy of the model in the case of asymptotically flat BHs with synchronised hair.

  
\section{The general framework}
\label{section2}

\subsection{ Komar integrals and Smarr relation}

We consider a general model in $D\geqslant 4$ 
spacetime dimensions,
consisting of Einstein's gravity  minimally 
coupled to some matter fields $\psi$ described by a Lagrangian density $\mathcal{L}_m$
\begin{eqnarray}
\label{action}
\mathcal{S}=\int d^Dx  \sqrt{-g}
\left[ \frac{R}{16\pi } +\mathcal{L}_m 
 \right]  \ ,
\end{eqnarray}
where $R$ is the spacetime Ricci scalar. Here and below we use geometrised units, setting Newton's constant and the speed of light to unity: $G=1=c$.

In this work we shall be  interested in stationary space-times with
$N$-azimuthal symmetries, where $N=1,2$, for $D=4,5$.
This implies the existence of $N + 1$ commuting Killing vectors,
$\xi \equiv \partial_t$, 
and $\eta^{(k)} \equiv \partial_{\vphi_k}$, for $k=1, \dots , N$.

Assuming asymptotic flatness,
the total (or ADM) mass $M$ and total angular momenta $J_{(k)}$ of the configurations
are obtained from Komar integrals~\cite{Komar:1958wp} (see also, $e.g.$~\cite{Townsend:1997ku}), at spatial infinity, associated with the corresponding Killing vector fields
\begin{equation}
M = -\frac{1}{16 \pi  } \frac{D-2}{D-3} \int_{S_{\infty}^{D-2}} \alpha \ , \qquad
J^{(k)} = \frac{1}{16 \pi  }  \int_{S_{\infty}^{D-2}} \beta^{(k)} \ ,
\end{equation}
with 
\begin{eqnarray}
\alpha_{\mu_1 \dots \mu_{D-2}} \equiv \epsilon_{\mu_1 \dots \mu_{D-2}
  \rho \sigma} \nabla^\rho \xi^\sigma \ ,
\qquad
\beta^{ (k) \mu_1 \dots \mu_{D-2}} \equiv \epsilon_{\mu_1 \dots \mu_{D-2}
  \rho \sigma} \nabla^\rho \eta^{(k) \sigma} \ .
\end{eqnarray}

We are mainly interested in BH solutions with a regular event horizon geometry (without any
restrictions on its topology, which for $D>4$ can be non-spherical~\cite{Emparan:2001wn,Emparan:2008eg,Kleihaus:2012xh}).
This horizon ${\cal H}$ has an associated (hyper)area of its spatial sections, $A_H$, and a temperature
$T_H$; there are also $N$ horizon angular velocities $ \Omega_{  H (k)} $
associated with the $N$-azimuthal symmetries.

Using Komar integrals computed at the event horizon, one  also defines a  horizon mass $M_{  H}$ and a set of $N$ horizon angular momenta
$J_{{   H}}^{ (k)}$,
\begin{eqnarray}
M_{  H} = -\frac{1}{16 \pi  } \frac{D-2}{D-3} \int_{{\cal H}} \alpha \ , \ \ \ 
J_{{  H} }^{(k)} = \frac{1}{16 \pi }  \int_{{\cal H}} \beta^{(k)} \ .
\end{eqnarray}
Then the following
 Smarr type mass formulae~\cite{Smarr:1972kt} hold: for the horizon quantities we have 
\begin{eqnarray}
\frac{D-3}{D-2} M_{   H} = \frac{1}{4  } T_H A_H + \sum_{(k)}  \Omega_{  H (k)} J_{{ H} }^{(k)},
\end{eqnarray}
whereas for the bulk quantities
\begin{eqnarray}
\label{smarr}
M = \frac{D-2}{D-3}
\bigg [
\frac{1}{4  } T_H A_H + \sum_{(k)} \Omega_{{  H} (k)} (J^{ (k)}-J_{(\psi)}^{ (k)})
\bigg ]+M_{(\psi)} \ .
\end{eqnarray}

In the above relations, $M_{(\psi)}$, $J_{(\psi)}^{ (k)}$ are the energy 
and angular momenta stored in the matter fields, with
\begin{eqnarray} 
M =  M_H+M_{(\psi)},~~~J^{ (k)}= J_{{ H} }^{(k)}+J_{(\psi)}^{ (k)} \ .
\label{split}
\end{eqnarray}
Via the Einstein equations, $M_{(\psi)}$ and $J_{(\psi)}^{ (k)}$
can be expressed as volume integrals for the appropriate components of the energy-momentum tensor (see $e.g.$~\cite{Townsend:1997ku}).

In addition to the above Smarr relations, the configuration should satisfy the first law of BH thermodynamics~\cite{Bardeen:1973gs},
\begin{eqnarray}
\label{1st}
dM=\frac{1}{4 G} T_H dA_H+\sum_{(k)} \Omega_{{  H} (k)} dJ^{ (k)}+ {\cal W},
\end{eqnarray}
where ${\cal W}$ denotes the work term(s) associated with the matter fields. In particular, for vacuum solutions, the following relation holds
\begin{eqnarray}
\label{first-law}
dM_H=\frac{1}{4} T_H dA_H +\sum_{(k)} \Omega_{{  H} (k)} dJ_{{ H} }^{(k)}~.
\end{eqnarray}

\subsection{The effective model}
We now turn into the assumptions of the effective model~\cite{Herdeiro:2017phl}, its statement and its corollaries. 

\bigskip

{\bf Assumption 1): Fiducial ``bald" BH}. 
One defines
a vacuum \textit{fiducial} BH solution\footnote{In $D=4$ the fiducial solution is obviously Kerr. But in higher dimensions, there can be different solutions for the same global charges and the horizon topology~\cite{Emparan:2001wn,Emparan:2008eg,Kleihaus:2012xh}, thus requiring the definition of the fiducial solution.}
which is approached smoothly as
 $M_{(\psi)}\to 0$,
${J_{(\psi)}^{ (k)}}\to 0$ 
($i.e$ with  the same symmetries
and horizon structure as the non-vacuum solution).
Moreover, at least in all cases discussed in this work, 
the horizon quantities of the fiducial BH  have known (in closed form) expressions in terms of the global charges (macroscopic degrees of freedom); schematically these are:
\begin{eqnarray}
\label{relation1}
A_H=A_H(M,J^{(k)}), \qquad T_H=T_H(M,J^{(k)}), \qquad \Omega_H^{(k)}=A_H(M,J^{(k)}) \ .
\end{eqnarray}

It will be useful to define a set of $N+1$ ``hairiness" parameters $\mathfrak{h}\equiv(p, q^{(k)})$, which measure the deviation of the hairy BH from the fiducial solution 
\begin{eqnarray}
\label{quantities2}
p\equiv  \frac{M_{(\psi)}}{M} \ , \qquad
q^{ (k)}\equiv \frac{J_{(\psi)}^{ (k)}}{J^{ (k)}} \ .
\end{eqnarray}

\bigskip

{\bf Assumption 2): ``Hair" matter field}. We assume that there is no work term associated with
the  matter fields in the 1st law (\ref{1st}). This assumption guarantees the matter is adding ``hair", without a Gauss law associated, and it will not introduce a different global charge (another macroscopic degree of freedom), that can be computed at the horizon, as, for instance, an electromagnetic ``matter" field would (electric charge). This still allows the ``hair" matter field to have a conserved Noether charge, see $e.g.$~\cite{Herdeiro:2015gia}, which is associated to a global, rather than gauge, symmetry.

\bigskip


{\bf Statement of the model}. The  horizon quantities 
${\cal Q}=(A_H,T_H,\Omega_H^{(k)})$
of the non-vacuum BH 
are still given by those of the corresponding fiducial BH, relation (\ref{relation1}),
expressed, however, in term of the horizon mass and angular momenta of the non-vacuum solution.
That is, one considers the substitution
\begin{eqnarray}
\label{subst}
{\cal Q}^{(\rm fid)}(M,J^{(k)}) \longrightarrow {\cal Q}^{(\rm HBH)}(M_H,J_H^{(k)}).
\end{eqnarray}

\bigskip

{\bf Corollary 1): Analytic formulas for horizon quantities.} From~\eqref{split} and (\ref{quantities2})
\begin{eqnarray}
\label{rels}
M_H=(1-p) M \ , \qquad J_{H}^{(k)}=(1-q^{(k)})J^{(k)} \ ;
\end{eqnarray} 
then, the horizon quantities of the hairy BH (\ref{subst}) can be expressed as
\begin{eqnarray}
\label{rel3s}
{\cal Q}^{(HBH)}(M,J^{(k)};\mathfrak{h}).
\end{eqnarray}
If the horizon quantities of the fiducial BH are expressed by analytic formulas ${\cal Q}^{(\rm fid)}(M,J^{(k)})$ so will be the horizon quantities of the hairy BH, eqs.~\eqref{rel3s}.

\bigskip

{\bf Corollary 2): Analytic relation between ``hairiness parameters".} 
The assumption that the horizon is still described by the vacuum reference BH, 
together with the 1st law 
(\ref{1st})
(without a work term ${\cal W}$)
implies that 
the matter fields satisfy the relation, for rotating BHs,\footnote{The case of static BHs is simpler and will be illustrated in Section~\ref{seccol}.}
\begin{eqnarray}
\label{1stn}
dM_{(\psi)}= \sum_{(k)} \Omega_{{  H} (k)} dJ_{(\psi)}^{(k)}~.
\end{eqnarray}
Then
we formally integrate the above relation
 treating $\Omega_{{  H} (k)} $ as a set of input parameter
(which is justified, since it belong to a different subsystem).
This results in
\begin{eqnarray}
\label{rel1}
M_{(\psi)}=  \sum_{(k)} \Omega_{{  H} (k)}J_{(\psi)}^{(k)},~~ {\rm or} ~
~M-M_{H}= \sum_{(k)}  \Omega_{{  H} (k)}(J ^{(k)}-J_{H}^{(k)}) \ .
\end{eqnarray}
Using $M_{(\psi)}=p M$, $J_{(\psi)}^{(k)}=q^{(k)} J^{(k)}$,
\eqref{rel1}  becomes
 \begin{eqnarray}
\label{rel2}
p = \sum_{(k)} \Omega_{{  H} (k)}  q^{(k)}  \frac{J^{(k)}}{M} \ .
\end{eqnarray}

{\bf Summary.} Relations (\ref{rel3s}) and (\ref{rel2}) are the central results of the proposed effective
model.
After considering them together,
 one can eliminate $p$ and 
 arrive at the following set of relations
\begin{eqnarray}
\label{relf}
A_H(M,J^{(k)};q^{(k)}), \qquad T_H(M,J^{(k)};q^{(k)}), \qquad \Omega_H^{(k)}(M,J^{(k)};q^{(k)}).
\end{eqnarray}
The explicit  form of these relations is case dependent; the approach, however, is general. These analytic equations can be compared with the numerical solution of the hairy BH to check the domain of validity of the effective model.

As a final remark, we observe that instead of eliminating $p$, eq. (\ref{rel2})
can be used to eliminate instead one of the parameters $q^{k}$,
which results in an equivalent form of (\ref{relf}). In practice, as usual in BH physics,  
it is natural to work in units set by
the BH mass ($i.e.$
with normalized quantities).

\section{Applications of the effective model}
\label{section30}

In this  section we shall consider specific applications of the effective model. The simplest such application is found for the following scalar matter content:
\begin{eqnarray}
\label{LS}
   \mathcal{L}_m=-  \left( \partial_a \Phi  \right)^\dagger \left( \partial^a \Phi \right)
 - U( \left| \Phi  \right|) 
 , 
\end{eqnarray}
where $\Phi$ is in general a scalar multiplet and $U( \left| \Phi  \right|) $ is a self-interactions potential.
The Einstein-Klein-Gordon equations possesses both solitons and hairy BH solutions. 
When the interaction potential includes a mass term, there are everywhere regular, asymptotically flat, stationary solitonic solutions known as boson stars~\cite{Schunck:2003kk}. Rotating boson stars~\cite{Yoshida:1997qf}, in particular, arise as a particular limit of Kerr BHs with (scalar) synchronised hair~\cite{Herdeiro:2014goa,Herdeiro:2015gia}. Both the solitonic rotating boson stars~\cite{Hartmann:2010pm} and the hairy BHs~\cite{Brihaye:2014nba,Herdeiro:2015kha} can be generalised to $D=5$.

In~\cite{Herdeiro:2017phl} the effective model was already applied to the simplest scalar~\cite{Herdeiro:2014goa} and vector BHs~\cite{Herdeiro:2016tmi} with synchronised hair.  In section~\ref{section3} we shall apply the effective model to the $D=4$ hairy BHs with self-interactions obtained in~\cite{Herdeiro:2015tia}. In section~\ref{section4} we shall apply it to the $D=5$ hairy BHs~\cite{Brihaye:2014nba}.

\subsection{$D=4$ BHs with synchronized scalar hair}
\label{section3}

\subsubsection{Predictions of the effective model}
Consider the $D=4$ Kerr BHs with synchronised hair~\cite{Herdeiro:2014goa,Herdeiro:2015gia,Herdeiro:2015tia} the reference solution is the Kerr metric~\cite{Kerr:1963ud}, for which $N=1$. Then the statement of the model, $cf.$ (\ref{subst}), is that the horizon quantities of the hairy BHs obey:
\begin{equation}
\label{tq1}
\Omega_H =\frac{M_H}{2J_H} \left(1-\chi \right), \ \ \ \ \ \ 
~
A_H=8\pi M_H^2 
\left(
1+\chi
\right), \ \ \ \ \ \ 
~
T_H=\frac{\chi}{4\pi M_H \left( 1+\chi \right)}, \qquad 
\chi\equiv \sqrt{1-\frac{J_H^2}{M_H^4}} \ .
\end{equation}
Next, we use (\ref{rels}), which in this case is simply, 
\begin{equation}
\label{split2}
M_H=(1-p)M \ , \qquad  J_H=J(1-q) \ ,
\end{equation}
to obtain the specific form of  (\ref{rel2}), which reads
\begin{eqnarray}
\label{si1}
 p M=\Omega_H J q \ .
\end{eqnarray}

We now introduce a set of reduced parameters, normalising the corresponding physical parameter by the ADM mass,
 \begin{equation}
\label{rq1}
j\equiv  \frac{J}{M^2}, \qquad 
a_H \equiv \frac{A_H}{16\pi M^2}, \qquad w_H\equiv \Omega_H M, \qquad t_H\equiv 8\pi M T_H \ .
\end{equation}
Then (\ref{si1}) takes the compact form
 \begin{equation} 
p=w_Hj q.
\label{pq1}
\end{equation}
Replacing~\eqref{tq1} in~\eqref{rq1}, making use of~\eqref{split2}  and choosing, via~\eqref{pq1}, $(p,w_H)$ as independent parameters, one arrives at the following expressions
\begin{eqnarray}
\label{expr1}
&&
q= p\frac{1+4(1-p)^2w_H^2}{p+4(1-p)^2w_H^2}\ , \qquad  
 j=\frac{p+4(1-p)^2w_H^2}{w_H[p+4(1-p)^2w_H^2]} \ , \\
 &&
  \label{expr4}
a_H= \frac{ (1-p)^2 }{1+4(1-p)^2w_H^2} \ , \qquad
t_H=\frac{1-4(1-p)^2w_H^2}{1-p} \ .
\end{eqnarray}
which can be taken as the predictions of the effective model, in the spirit of eq.~\eqref{relf}. It is also possible to show that
\begin{eqnarray}
\label{rs3}
\frac{p}{q}=1-\frac{a_H}{1-p}  \ ,
\end{eqnarray}
which implies, according to the effective model, that  $p<q$ for a hairy BH.

Finally, we observe that
it is possible to express all quantities solely in terms of the hairiness parameters $(p,q)$:
\begin{eqnarray}
\label{rj1}
w_H=\frac{1}{2(1-p)}\sqrt{\frac{p(1-q)}{q-p}},~~  j=\frac{2(1-p)\frac{p}{q}}{\sqrt{\frac{p}{q}\frac{1-q}{1-\frac{p}{q}}}},~~ 
a_H=(1-p)\left(1-\frac{p}{q}\right), ~~  t_H=\frac{1-\frac{p}{q}(2-q)}{(1-p)\left(1-\frac{p}{q}\right)} ~.
\end{eqnarray}

\subsubsection{Validating the effective model}
For $D=4$, we restrict our study here to the simplest case of a scalar singlet (single complex field), with
\begin{eqnarray}
\label{scalar4}
\Phi=\phi(r,\theta)e^{i(m \varphi-w t)} \ .
\end{eqnarray} 
We will also focus on the scalar field potential  \cite{Herdeiro:2015tia}
\begin{eqnarray}
\label{pot}
 U(|\Phi|)= \mu^2\left|\phi\right|^2 + \lambda\left|\phi\right|^4,
\end{eqnarray}  
where  $\mu$ is the scalar field mass and the self-coupling is positive, $ \lambda>0$. 
 This complements the study in~\cite{Herdeiro:2017phl} which addresses the case $\lambda=0$. 

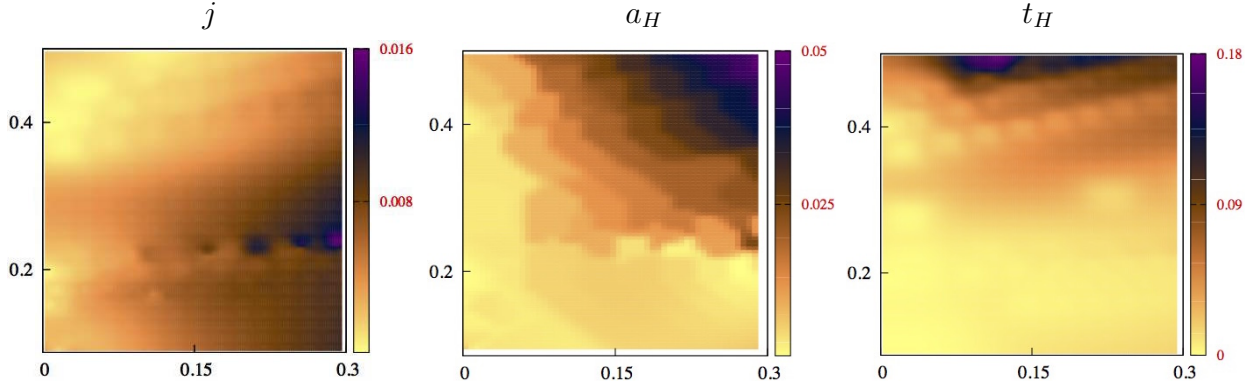
\begin{figure}[ht]
\begin{center}
\includegraphics[width=0.33\textwidth]{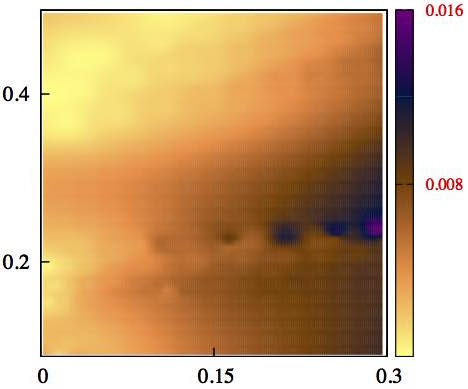}
\includegraphics[width=0.33\textwidth]{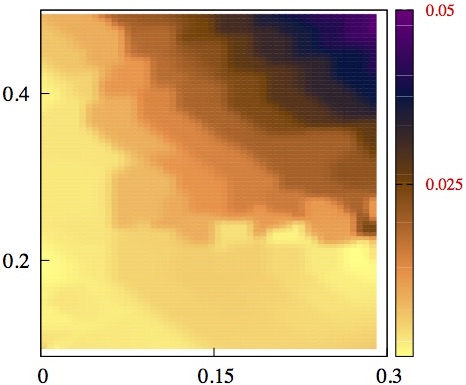}
\includegraphics[width=0.32\textwidth]{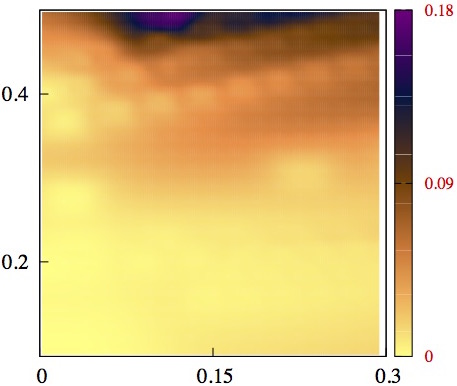}
\begin{picture}(0,0)
\put(150,150){$t_H$}
\put(0,150){$a_H$}
\put(-160,150){$ j$}
\end{picture}
\end{center}
\caption{\small 
The relative errors are shown in a strip on the $(p,w_H)$-plane for
reduced angular momentum $j$ (left panel),  
hairiness parameter $q$ (center)
and 
reduced temperature $t_H$ (right panel) 
for $D=4$ BHs with self-interacting scalar hair, with $\lambda=100$. The Kerr limit ($p=0$) is the left vertical axis.
}
\label{d4errors}
\end{figure}

In Fig.~\ref{d4errors} we exhibit the relative errors
$|1-{\cal Q}^{(\rm model)}/{\cal Q}^{(\rm num)}|$  for the quantities ${\cal Q}=(j,q,t_H)$ in terms of the parameters $(p,w_H)$. Observe that the self-interactions only affect ${\cal Q}^{(\rm num)}$; the effective model prediction ${\cal Q}^{(\rm model)}$ is insensitive to the self-interactions. The overall errors in this domain are comparable to those in~\cite{Herdeiro:2017phl}, which is somewhat expected because the effect of the self interactions is small (for the numerical solutions) in the region studied, which is in the vicinity of the existence line $(p=0=q)$, wherein the hairy BHs reduce to Kerr BHs. 
Remarkably, even for fairly large values of $p$, such as $p\sim 0.3$, the effective models gives an error of only a few percent, for low $\omega_H$. For the reduced angular momentum the error is of $\sim 1\%$ within the whole domain $p\in [0,0.3]$, $\omega_H\in[0,0.5]$!

\subsection{$D=5$ BHs with synchronized scalar hair}
\label{section4}
We now consider Myers-Perry BHs with synchronised hair~\cite{Brihaye:2014nba,Herdeiro:2015kha}. $D=5$ rotating BHs with scalar hair possess, generically, two independent angular momenta and may even have a more general topology of the event horizon~\cite{Herdeiro:2017oyt}. Here, we shall focus on the case with two equal angular momenta~\cite{Brihaye:2014nba}. In $D=5$ a qualitative difference with respect to the $D=4$ case is that there is a mass gap between the hairy BHs and the fiducial model, which is taken to be the Myers-Perry vacuum solution~\cite{Myers:1986un}, with two equal angular momenta. 

\subsubsection{Predictions of the effective model}
$D=5$ BHs with two equal-magnitude angular momenta have
\begin{eqnarray}
J^1=J^2\equiv J \ , \qquad \Omega_{H(1)}=\Omega_{H(2)}\equiv \Omega_{H} \ .
\end{eqnarray}
We also consider only the case with a spherical horizon topology.
For such solutions the isometry group is enhanced from $\mathbb{R}_t \times U(1)^{2}$
to $\mathbb{R}_t \times U(2)$, where $\mathbb{R}_t$ refers to time translations.
  This symmetry enhancement allows to factorize the angular dependence
and thus leads to ordinary differential equations (not partial differential equations).

The fiducial solution  is, in this case, the double spinning Myers-Perry BH.
Then the statement of the model, $cf.$ (\ref{subst}), is that the horizon quantities of the hairy BHs obey:
\begin{equation}
\label{i1}
\Omega_H =\frac{M_H}{3J_H} \left(1- \bar{\chi}   \right),
 \ \  
A_H=\frac{16}{3}\sqrt{\frac{2\pi}{3}} M_H^{3/2} 
 \left(1+\bar{\chi}\right),
 \ \  
T_H=\sqrt{\frac{3}{8\pi M_H}}\frac{\bar{\chi}}
{ 1+\bar{\chi}}, 
\ \ \ \bar{\chi}\equiv \sqrt{1-\frac{27 \pi J_H^2}{8 M_H^3}} .
\end{equation}

Again, it is convenient to define quantities normalized $w.r.t.$ the ADM mass of the BHs.
The $D=5$ usual conventions in the literature are
\begin{eqnarray}
\label{quantities1}
j= \frac{3}{2}\sqrt{\frac{3\pi}{2}}\frac{J}{M^{3/2}},~~
a_H=\frac{3}{32}\sqrt{\frac{3}{2\pi}}\frac{A_H}{ M^{3/2}},~~
w_H=\sqrt{\frac{8}{3\pi}}\Omega_H \sqrt{M},~~
t_H=4\sqrt{\frac{2\pi}{3}}T_H \sqrt{M}~.
\end{eqnarray}
In the absence of hair, $i.e.$ for $p=q=0$, corresponding to a Myers-Perry BH, the following relations hold
\begin{eqnarray}
\label{Kerr1}
j=\frac{2w_H}{1+w_H^2}, \qquad a_H=\frac{1}{1+w_H^2}, \qquad t_H=1-w_H^2,
\end{eqnarray}
with $
0\leqslant w_H \leqslant 1; 
$
the limits correspond, respectively, to the Schwarzschild-Tangherlini~\cite{Tangherlini:1963bw}  and extremal Myers-Perry BHs. 
 
Repeating the procedure described in the previous section, choosing again $(p,w_H)$ as the independent parameters, yields the following simple expressions
\begin{eqnarray}
\label{expr1b}
&&
q= 3p\frac{1+(1-p) w_H^2}{3p+(4-p)(1-p)w_H^2}, \qquad 
 j=\frac{3p+(4-p)(1-p)w_H^2}{2w_H(1+(1-p)w_H^2)}, \\
&&
a_H= \frac{ (1-p)^{3/2} }{1+(1-p) w_H^2}, \qquad \qquad \qquad
 \label{expr4b}
t_H= \frac{1-(1-p) w_H^2}{\sqrt{1-p}}. 
\end{eqnarray} 
Again, these can be regarded as the predictions of the effective model, in the spirit of eq.~\eqref{relf}.

\subsubsection{Validating the effective model}
\label{Ansatz}

The matter content  in this case, consistent with the aforementioned symmetries is found by taking
  $\Phi$ a complex doublet scalar field   \cite{Hartmann:2010pm}, with
\begin{equation}
 \Phi = \phi(r) e^{-i \omega  t} 
 \left( \begin{array}{c} 
   \sin\theta  e^{i \vphi_1} \\ \cos\theta  e^{i \vphi_2} 
 \end{array} \right) 
 , \label{phi} 
\end{equation} 
where $\theta  \in [0,\pi/2]$, $(\varphi_1,\varphi_2) \in [0,2\pi]$, and $t$ denotes the time coordinate.
Concerning the scalar field potential, we restrict our study to the simplest case with
   \begin{eqnarray}
   U(|\Phi|) =\mu^2 \Phi^\dagger \Phi  =\mu^2 \phi(r)^2
   \end{eqnarray}
where $\mu$ corresponds to the scalar field mass.

The corresponding $D=5$ hairy BHs are discussed in
\cite{Brihaye:2014nba}. 
Likewise their four dimensional counterparts, the solutions are supported by rotation and have no static limit.
The main difference with respect to the $D=4$ case if the absence of the existence line.
That is, in the limit of vanishing Noether charge density, the scalar field becomes point-wise arbitrarily small and the
geometry becomes, locally, arbitrarily close to that of a specific set of Myers-Perry BHs.
There remains, however, a global difference with respect to the latter, manifest in a finite mass gap.
Thus the hair of these $D=5$ hairy BHs is intrinsically non-linear.

We have found that the effective model provides an accurate description of the hairy BHs in the vicinity of the ``marginally bound set".
This is the natural $D=5$ counterpart of the $D=4$ existence line, wherein the matter field becomes point-wise arbitrarily small, even though the global charges do not not vanish. This line is approached for $w\to \mu$.

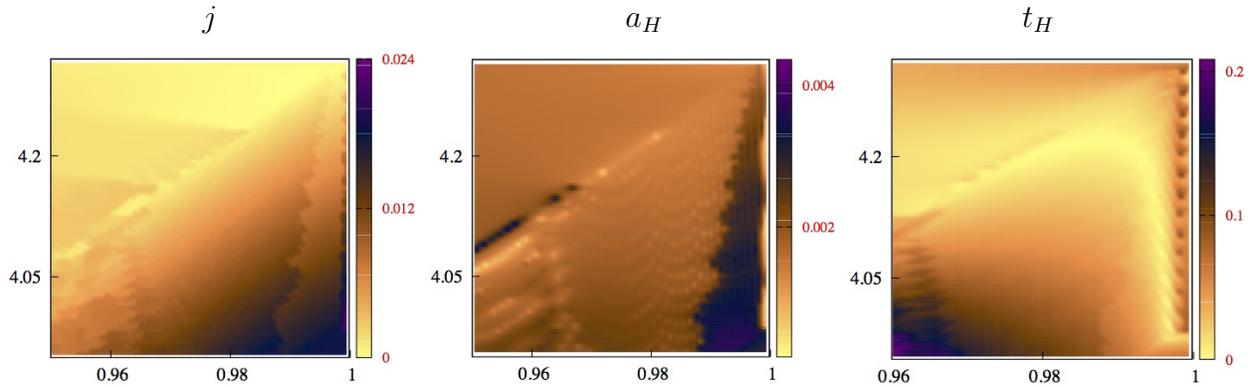
\begin{figure}[h!]
\begin{center}
\includegraphics[width=0.33\textwidth]{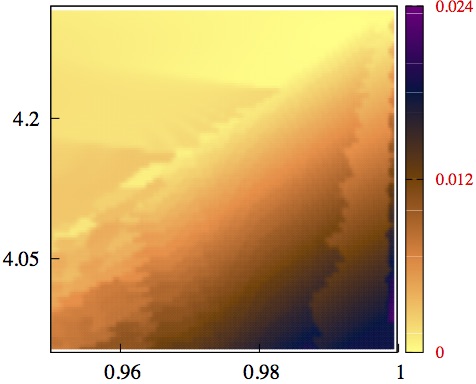}
\includegraphics[width=0.33\textwidth]{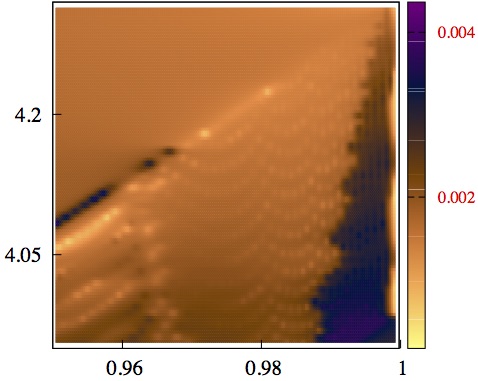}
\includegraphics[width=0.32\textwidth]{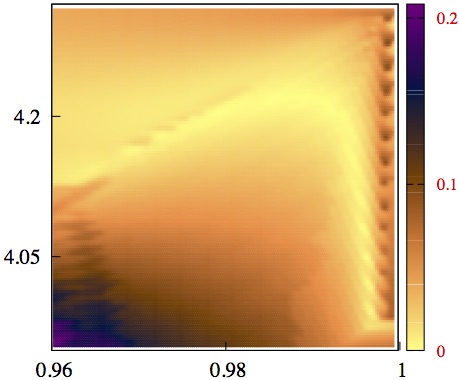}
\begin{picture}(0,0)
\put(150,150){$t_H$}
\put(0,150){$a_H$}
\put(-160,150){$ j$}
\end{picture}
\end{center}
\caption{\small 
The relative errors are shown in a strip on the $(p,w_H)$-plane for
reduced angular momentum $j$ (left panel),  
hairiness parameter $q$ (center)
and 
reduced temperature $t_H$ (right panel) 
for $D=5$ BHs with scalar hair. 
}
\label{d5errors}
\end{figure}

In Fig.~\ref{d5errors} we display the relative errors of the effective model for Myers-Perry BHs with synchronised hair. As before, these relative errors are
$|1-{\cal Q}^{(\rm model)}/{\cal Q}^{(\rm num)}|$, and are exhibited for the quantities ${\cal Q}=(j,q,t_H)$ in terms of the parameters $(p,w_H)$.
Impressively, even for extremely large values of $p$, such as $p\sim 0.98$, the effective model gives an error of less than one percent for $a_H$, for the $\omega_H$ range plotted! For $j$ the error reaches $\sim \mathcal{O}(2\%)$ whereas for $t_H$ is one order of magnitude larger $\sim \mathcal{O}(20\%)$. But even in the latter case it is considerably smaller than the naive expectation that the error should be of order of the deviation from the fiducial BH. As already mentioned, these hairy BHs do not continuously connect globally to the fiducial BH -- there is always a mass gap~\cite{Brihaye:2014nba}, the minimum value of $p$ for the hairy solutions being roughly $p\sim 0.93$.

\section{Limitations of the effective model}
\label{section5}

\subsection{$D=5$ $AdS$ BHs with synchronized scalar hair}

The first example of BHs with synchronized hair in the literature are obtained in $D=5$ $AdS$ asymptotics~\cite{Dias:2011at}. It is interesting to inquire if the effective model may still provide a useful description in that case as well, which is qualitatively different due to the $AdS$ asymptotics. 

To address this question, we have performed an independent investigation of the hairy BHs in~\cite{Dias:2011at}. 
These solutions can be studied within the same framework in~\cite{Brihaye:2014nba}, already mentioned in Section \ref{Ansatz}.
The presence of a mass term, however, is not necessary in this case, since the AdS asymptotics provides the necessary confining mechanism. Thus, following~\cite{Dias:2011at}, we set $\mu=0$  in what follows.

\begin{figure}[h!]
\hbox to\linewidth{\hss%
	\resizebox{8cm}{6cm}{\includegraphics{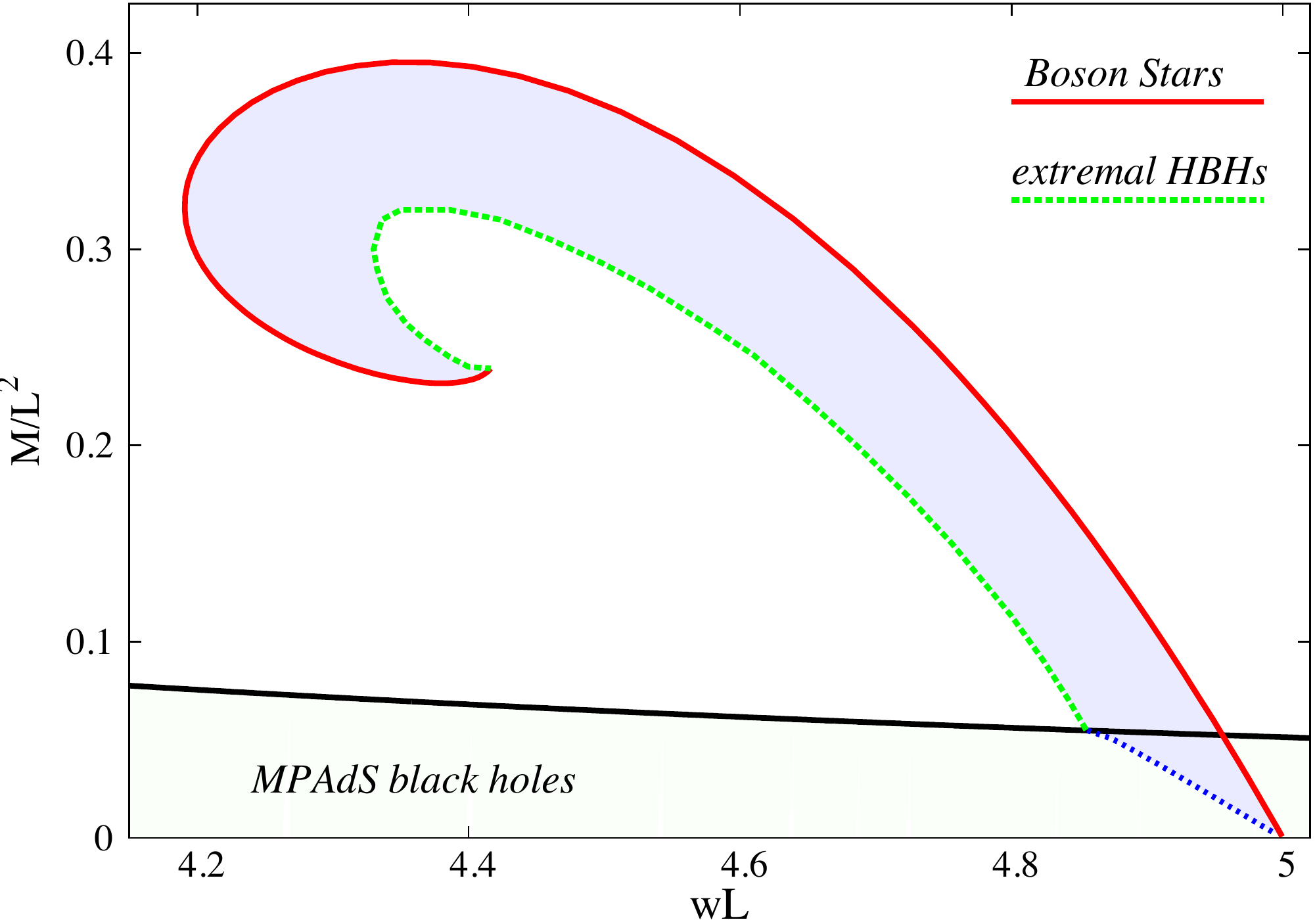}}
\hss}
\caption{\small  
The domain of existence in $(w,M)$-plane is shown for BHs with synchronised hairy with equal angular momenta, in AdS$_5$.
}
\label{AdS}
\end{figure}

The domain of existence of the solutions is shown in Fig.~\ref{AdS}. This plot manifests the striking analogy with the $D=4$ asymptotically flat case. The domain of hairy BHs is bounded by a solitonic limit (red solid line - corresponding to $AdS$ rotating boson stars, with equal angular momenta~\cite{Hartmann:2010pm} ), by the ``bald limit", defining the \textit{existence line} (blue dotted line - corresponding to equal angular momenta Myers-Perry $AdS$ BHs~\cite{Hawking:1998kw}) and a set of extremal ($i.e.$ zero temperature, extremal hairy BHs).

 {\small \hspace*{3.cm}{\it  } }

Concerning the validity of the effective model, we shall now discuss how it holds only partially. Following the procedure in the previous sections, we first identify the fiducial BH has the equal angular momentum, Myers-Perry-$AdS$ BH~\cite{Hawking:1998kw}.  Then, the statement of the model, $cf.$ (\ref{subst}), is that the horizon quantities of the hairy BHs obey the thermodynamical relations of the fiducial BH, in terms of the macroscopic degrees of freedom measured at the horizon. Recall that the horizon quantities $(A_H,T_H, \Omega_H)$ of the fiducial BH are
\begin{equation}
A_H=\frac{2\pi^2 L^3x^4}{U}, \qquad 
T_H=\frac{1}{2\pi L U} \left[1+2x^2-\frac{2a^2}{L^2x^2}(1+x^2)^2 \right], \qquad
\Omega_H=\frac{a}{L^2}\left(1+\frac{1}{x^2}\right).
\end{equation}
where $L$ is the $AdS$ radius and
\begin{eqnarray}
U\equiv \sqrt{x^2\left(1-\frac{a^2}{L^2}\right)-\frac{a^2}{L^2}}.
\end{eqnarray}
$x$ and $a$ are the two parameters of the solution which  determine the horizon mass and angular momentum, according to:
\begin{eqnarray}
M_H=\frac{3\pi L^2}{8}\frac{x^4(1+2x^2)}{U^2}, \qquad
J_H=\frac{\pi a L^2}{4}\frac{x^4(1+x^2)}{U^2}
\end{eqnarray}

\begin{figure}[ht!]
\begin{center}
{\label{c1}\includegraphics[width=9cm]{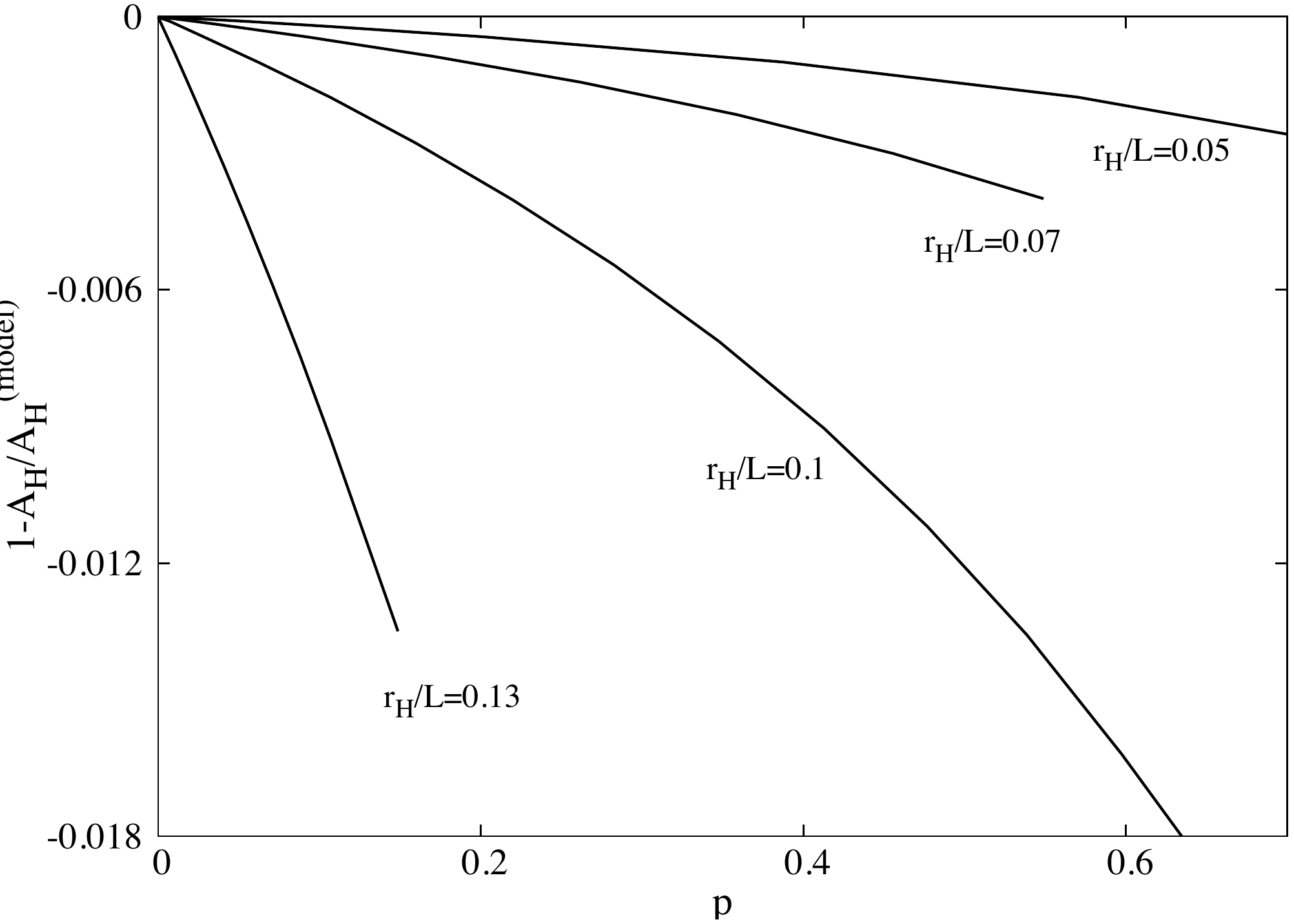}}
\caption{The relative errors are shown for the event horizon area as a function
of the  
hairiness parameter $p$, for several values  of the event horizon radius. 
\label{errors-AdS-AH}
}
\end{center}
\end{figure}

Our study reveals that, in a region close to the existence line, the hairy BHs  are still well described by the effective model. This can be confirmed in Fig. \ref{errors-AdS-AH} (left panel) where we show the relative errors for the horizon area for several values of the event horizon radius $r_h$, comparing the value computed from the effective model, $A_H^{(\rm model)}$, with the one computed from the numerical solutions $A_H^{(\rm num)}$. Similar results were found for $T_H$ and $\Omega_H$.

This case, however, makes clear a limitation in the applicability of the model. In fact, a crucial ingredient of the formalism is missing in the $AdS$ case.
Although one can formally define the hairiness parameters
$p,q$ (where $M_\psi$, $J_\psi$ are compute as volume integrals),
the splitting of the total mass and and angular momentum as the sum of the horizon charges plus 
the contribution from the matter fields is not possible in the presence of a cosmological constant\footnote{One can
easily verify that $M\neq M_H$ even for a vacuum Schwarzschild $AdS$ BH.}
\begin{eqnarray}
M\neq M_H+M_\psi,~~~J \neq J_H+J_\psi.
\end{eqnarray}
Thus $A_H,T_H,\Omega_H$
cannot be further re-expressed in terms of $(p,q)$
and the effective description holds only $locally$,
at the horizon level.

\subsection{$D=4$ coloured BHs}
\label{seccol}

The first clear counterexamples to  the no-hair conjecture was found in 1989 in Einstein-Yang-Mills (EYM) theory~\cite{Volkov:1989fi}, dubbed \textit{coloured BHs}, in the wake of the discovery, 
by Bartnik and  Mckinnon, that solitonic horizonless
configurations exist in the same model~\cite{Bartnik:1988am}.
This contrasts with the case of Einstein-Maxwell theory,
wherein the absence of solitons is rigorously established~\cite{Shiromizu:2012hb}. Coloured BHs are hairy since they have a non-trivial matter configuration outside their regular event horizon, and the solutions are 
no longer completely determined by their global charges. These BHs are unstable \cite{Zhou:1991nu} already in the 
spherically symmetric case.
A (rather old)  review of these solutions 
can be found in~\cite{Volkov:1998cc}.

The EYM- SU(2) ``matter" Lagrangian density is
\begin{eqnarray}
\label{action}
\mathcal{L}_m=
-\frac{1}{2 }{\rm Tr}\{F_{\mu \nu }F^{\mu \nu} \} \ ,
\end{eqnarray}
where
$F_{\mu\nu}$ is
 the
 field strength tensor $F_{\mu\nu} = \partial_\mu A_\nu - \partial_\nu A_\mu + i e  [A_\mu, A_\nu]$,  $e$ is the gauge coupling constant,  the gauge potential is $A_\mu= \tau_a A_\mu^a/2$, and $\tau_a$ are the Pauli matrices.
Here, $\mu,\nu$ are space-time indices running from 1 to 4 and the gauge index $a$ is running from 1 to 3.

The static, spherically symmetric EYM hairy BHs possess a  purely magnetic YM field, with
a single gauge potential $w(r)$.  The magnetic flux at infinity vanishes and, as a result, there is a single global charge -- the ADM mass $M$.
These BHs consist of a 1-discrete parameter family of solutions, labelled by the integer $k \geqslant 1$,
 which denotes the node number of the function $w(r)$. In the following discussion we focus on the solutions with $k=1$.

The domain of existence of this solutions can be qualitatively described as follows. 
Firstly, no upper bound exists for the horizon size.  In the large horizon size limit, the Einstein equations decouple from the gauge sector, yielding a Yang-Mills system on a fixed Schwarzschild BH background. In this decoupling limi there is a known exact solution (in closed form)~\cite{BoutalebJoutei:1979va}.
As the horizon size shrinks to zero, the Bartnik -Mckinnon family of solitons is recovered. 
Further details  can be found in \cite{Volkov:1998cc}.
 
\begin{figure}[h!]
\begin{center}
\includegraphics[width=0.55\textwidth]{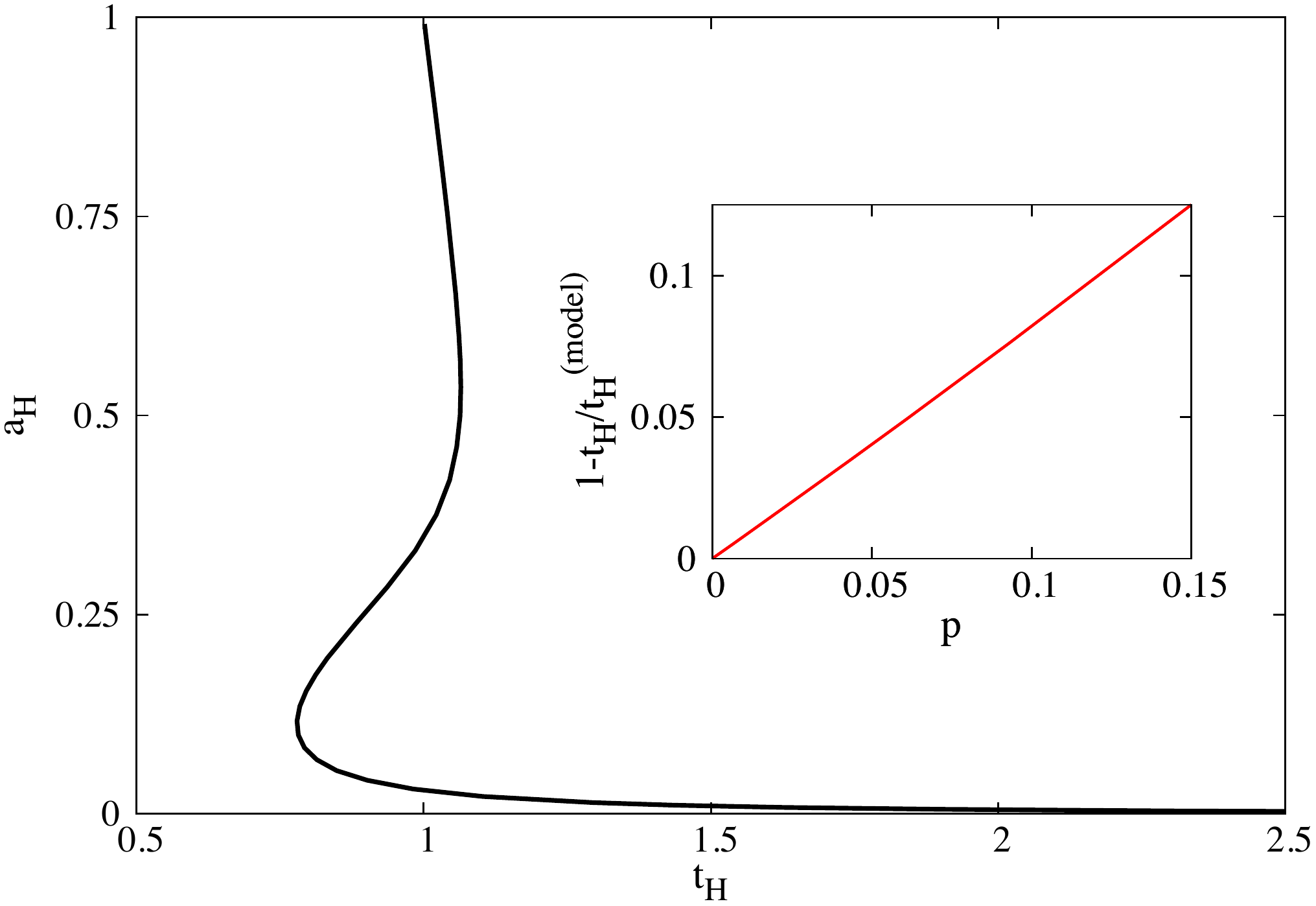} 
\caption{  The (reduce) temperature-area diagram for EYM BHs is exhibited together with the relative errors for the temperature.
}
\label{figYM}
\end{center}
\end{figure}

For applying the effective model, the obvious fiducial metric is that of a Schwarzschild BH.
Then, applying the formalism in Section~\ref{section2} leads to the following predictions
\begin{eqnarray}
\label{EYM-res}
a_H=(1-p)^2, \qquad t_H=\frac{1}{1-p}.
\end{eqnarray}
Computing the relative errors, like before, for instance of the temperature one concludes the error is of the order of $p$ - Fig.~\ref{figYM}. This the unremarkable result one expects in general: that the error in the effective model is of the order of the deviation from the bald BH. Thus, in this case, the effective model is not particularly accurate.

\section{Discussion and final remarks}
\label{section6}
	
In this paper we have discussed an effective model for computing, analytically, several quantities for hairy BHs, in terms of the corresponding quantities of a fiducial bald BH. This model, already considered in~\cite{Herdeiro:2017phl} for Kerr BHs with synchronised scalar~\cite{Herdeiro:2014goa} and vector~\cite{Herdeiro:2016tmi}  hair, was applied  to two other examples of BHs with synchronised hair: the $D=4$ BHs with scalar self-interactions~\cite{Herdeiro:2015tia} and to the   $D=5$ ``hairy" Myers-Perry BHs~\cite{Brihaye:2014nba} in section~\ref{section4}. In all these cases one finds that the relative errors in the quantities provided by the model can be considerably smaller than those one would naively expect, namely that these errors should be of the order of deviation from the fiducial solution. 

To illustrate the limitations of the model, we have also considered two other examples in Section~\ref{section5}: the $D=5$ $AdS$ Myers-Perry BHs with synchronised hair~\cite{Dias:2011at} and to the coloured BHs in Einstein-Yang-Mills theory~\cite{Volkov:1989fi}. The former case shows that one step of the model, namely the splitting between horizon and ADM quantities may be subtle in non-asymptotically flat spacetimes; still one may use the model and find small errors, as in the asymptotically flat case. The latter case illustrates that the model may give errors of the order of the deviations from the fiducial BH, thus unimpressive. 

In this set of applications, the example of coloured BHs is the exceptional one. All the remaining examples rely on the synchronisation mechanism to endow rotating BHs with hair. So, why is this effective model performing better for this sort of ``hairy" BHs? A possible answer is that, in these cases, there is a separation of scales. The hair field has its largest amplitude not at the horizon but at some distance thereof (see $e.g.$~\cite{Herdeiro:2015gia}) -- defining a different scale from that of the horizon --, unlike the coloured case, where it decreases away from the  horizon. This suggest a more efficient decoupling between the two sub-systems (the ``bald" horizon and the matter ``hair") occurs in this case, allowing the horizon to remain fiducial BH-like even for larger amplitudes of the matter field. It would be interesting to further explore this suggestion.

\section*{Acknowledgements}

C. H. and E.R. acknowledges funding from the FCT-IF programme.  This work was supported by the European  Union's  Horizon  2020  research  and  innovation  programme  under  the H2020-MSCA-RISE-2015 Grant No.   StronGrHEP-690904, the H2020-MSCA-RISE-2017 Grant No. FunFiCO-777740  and  by  the  CIDMA  project UID/MAT/04106/2013. The authors  would also  like  to  acknowledge networking support by the COST Action GWverse CA16104.


	

\begin{thebibliography}{99}
\begin{small} 

\bibitem{Chrusciel:2012jk}
  P.~T.~Chrusciel, J.~Lopes Costa and M.~Heusler,
  Living Rev.\ Rel.\  {\bf 15} (2012) 7
  [arXiv:1205.6112 [gr-qc]].

\bibitem{Bekenstein:1973ur}
  J.~D.~Bekenstein,
  Phys.\ Rev.\ D {\bf 7} (1973) 2333.

\bibitem{Hawking:1974sw}
  S.~W.~Hawking,
  Commun.\ Math.\ Phys.\  {\bf 43} (1975) 199
   Erratum: [Commun.\ Math.\ Phys.\  {\bf 46} (1976) 206].

\bibitem{Strominger:1996sh}
  A.~Strominger and C.~Vafa,
  Phys.\ Lett.\ B {\bf 379} (1996) 99
  [hep-th/9601029].

\bibitem{Callan:1996dv}
  C.~G.~Callan and J.~M.~Maldacena,
  Nucl.\ Phys.\ B {\bf 472} (1996) 591
  doi:10.1016/0550-3213(96)00225-8
  [
  
\bibitem{Ruffini:1971bza}
  R.~Ruffini and J.~A.~Wheeler,
  Phys.\ Today {\bf 24} (1971) no.1,  30.

\bibitem{Komar:1958wp}
  A.~Komar,
  Phys.\ Rev.\  {\bf 113} (1959) 934.

\bibitem{Bekenstein:1996pn}
  J.~D.~Bekenstein,
  In *Moscow 1996, 2nd International A.D. Sakharov Conference on physics* 216-219
  [gr-qc/9605059].

\bibitem{Herdeiro:2015waa}
  C.~A.~R.~Herdeiro and E.~Radu,
  Int.\ J.\ Mod.\ Phys.\ D {\bf 24} (2015) no.09,  1542014
  [arXiv:1504.08209 [gr-qc]].

\bibitem{Volkov:2016ehx}
  M.~S.~Volkov,
  arXiv:1601.08230 [gr-qc].
  
\bibitem{Herdeiro:2017phl}
  C.~A.~R.~Herdeiro and E.~Radu,
  Phys.\ Rev.\ Lett.\  {\bf 119} (2017) no.26,  261101
  [arXiv:1706.06597 [gr-qc]].
  
\bibitem{East:2017ovw}
  W.~E.~East and F.~Pretorius,
  Phys.\ Rev.\ Lett.\  {\bf 119} (2017) no.4,  041101
  [arXiv:1704.04791 [gr-qc]].
  
\bibitem{Herdeiro:2015tia}
  C.~A.~R.~Herdeiro, E.~Radu and H.~Rœnarsson,
  Phys.\ Rev.\ D {\bf 92} (2015) no.8,  084059
  [arXiv:1509.02923 [gr-qc]].
  
\bibitem{Brihaye:2014nba}
  Y.~Brihaye, C.~Herdeiro and E.~Radu,
  Phys.\ Lett.\ B {\bf 739} (2014) 1
  [arXiv:1408.5581 [gr-qc]].

\bibitem{Dias:2011at}
  O.~J.~C.~Dias, G.~T.~Horowitz and J.~E.~Santos,
  JHEP {\bf 1107} (2011) 115
  [arXiv:1105.4167 [hep-th]].
  
\bibitem{Volkov:1989fi}
  M.~S.~Volkov and D.~V.~Galtsov,
  JETP Lett.\  {\bf 50} (1989) 346
   [Pisma Zh.\ Eksp.\ Teor.\ Fiz.\  {\bf 50} (1989) 312];
\\
  H.~P.~Kuenzle and A.~K.~M.~Masood- ul- Alam,
  J.\ Math.\ Phys.\  {\bf 31} (1990) 928;
 \\
  P.~Bizon,
  Phys.\ Rev.\ Lett.\  {\bf 64} (1990) 2844.

\bibitem{Townsend:1997ku}
  P.~K.~Townsend,
  gr-qc/9707012.

\bibitem{Emparan:2001wn}
  R.~Emparan and H.~S.~Reall,
  Phys.\ Rev.\ Lett.\  {\bf 88} (2002) 101101
  [hep-th/0110260].
  
\bibitem{Emparan:2008eg}
  R.~Emparan and H.~S.~Reall,
  Living Rev.\ Rel.\  {\bf 11} (2008) 6
  [arXiv:0801.3471 [hep-th]].

\bibitem{Kleihaus:2012xh}
  B.~Kleihaus, J.~Kunz and E.~Radu,
  Phys.\ Lett.\ B {\bf 718} (2013) 1073
  [arXiv:1205.5437 [hep-th]].

\bibitem{Smarr:1972kt}
  L.~Smarr,
  Phys.\ Rev.\ Lett.\  {\bf 30} (1973) 71
   Erratum: [Phys.\ Rev.\ Lett.\  {\bf 30} (1973) 521].

\bibitem{Bardeen:1973gs}
  J.~M.~Bardeen, B.~Carter and S.~W.~Hawking,
  Commun.\ Math.\ Phys.\  {\bf 31} (1973) 161.

\bibitem{Schunck:2003kk}
  F.~E.~Schunck and E.~W.~Mielke,
  Class.\ Quant.\ Grav.\  {\bf 20} (2003) R301
  [arXiv:0801.0307 [astro-ph]].

\bibitem{Yoshida:1997qf}
  S.~Yoshida and Y.~Eriguchi,
  Phys.\ Rev.\ D {\bf 56} (1997) 762.

\bibitem{Herdeiro:2014goa}
  C.~A.~R.~Herdeiro and E.~Radu,
  Phys.\ Rev.\ Lett.\  {\bf 112} (2014) 221101
  [arXiv:1403.2757 [gr-qc]].

\bibitem{Herdeiro:2015gia}
  C.~Herdeiro and E.~Radu,
  Class.\ Quant.\ Grav.\  {\bf 32} (2015) no.14,  144001
  [arXiv:1501.04319 [gr-qc]].

\bibitem{Hartmann:2010pm}
  B.~Hartmann, B.~Kleihaus, J.~Kunz and M.~List,
  Phys.\ Rev.\ D {\bf 82} (2010) 084022
  [arXiv:1008.3137 [gr-qc]].
 

\bibitem{Herdeiro:2015kha}
  C.~Herdeiro, J.~Kunz, E.~Radu and B.~Subagyo,
  Phys.\ Lett.\ B {\bf 748} (2015) 30
  [arXiv:1505.02407 [gr-qc]].

\bibitem{Herdeiro:2016tmi}
  C.~Herdeiro, E.~Radu and H.~Runarsson,
  Class.\ Quant.\ Grav.\  {\bf 33} (2016) no.15,  154001
  [arXiv:1603.02687 [gr-qc]].

\bibitem{Kerr:1963ud}
  R.~P.~Kerr,
  Phys.\ Rev.\ Lett.\  {\bf 11} (1963) 237.

\bibitem{Herdeiro:2017oyt}
  C.~Herdeiro, J.~Kunz, E.~Radu and B.~Subagyo,
  Phys.\ Lett.\ B {\bf 779} (2018) 151
  [arXiv:1712.04286 [gr-qc]].

\bibitem{Myers:1986un}
  R.~C.~Myers and M.~J.~Perry,
  Annals Phys.\  {\bf 172} (1986) 304.

\bibitem{Tangherlini:1963bw}
  F.~R.~Tangherlini,
  Nuovo Cim.\  {\bf 27} (1963) 636.

\bibitem{COLSYS}
 U. Ascher, J. Christiansen, R.~D. Russell,
 Mathematics of Computation 33 (1979) 659;
 ACM Transactions 7 (1981) 209.  
  
\bibitem{Hawking:1998kw}
  S.~W.~Hawking, C.~J.~Hunter and M.~Taylor,
  Phys.\ Rev.\ D {\bf 59} (1999) 064005
  [hep-th/9811056].

\bibitem{Bartnik:1988am}
  R.~Bartnik and J.~Mckinnon,
  Phys.\ Rev.\ Lett.\  {\bf 61} (1988) 141.
  
\bibitem{Kleihaus:1997mn}
  B.~Kleihaus and J.~Kunz,
  Phys.\ Rev.\ D {\bf 57} (1998) 834
  doi:10.1103/PhysRevD.57.834
  [gr-qc/9707045].

\bibitem{BoutalebJoutei:1979va}
  H.~Boutaleb-Joutei, A.~Chakrabarti and A.~Comtet,
  Phys.\ Rev.\ D {\bf 20} (1979) 1884.
\bibitem{Shiromizu:2012hb}
  T.~Shiromizu, S.~Ohashi and R.~Suzuki,
  Phys.\ Rev.\ D {\bf 86} (2012) 064041
  [arXiv:1207.7250 [gr-qc]].


\bibitem{Zhou:1991nu}
  Z.~h.~Zhou and N.~Straumann,
  Nucl.\ Phys.\ B {\bf 360} (1991) 180.
  
\bibitem{Volkov:1998cc}
  M.~S.~Volkov and D.~V.~Gal'tsov,
  Phys.\ Rept.\  {\bf 319} (1999) 1
  [hep-th/9810070].
   
  


	


	


	

\end{small}
\end{thebibliography}
\end{document}